\documentclass[11pt]{article}
\usepackage{moriond,epsfig}
\usepackage[latin1]{inputenc}
\usepackage[OT1]{fontenc}

\def\beq{\begin{equation}}
\def\eeq{\end{equation}}
\def\dde{{\rm DE}}

\def\PRL{{\em Phys. Rev. Lett.~}}
\def\PRD{{\em Phys. Rev.} D}

\def\be{\begin{equation}}
\def\ee{\end{equation}}
\def\bea{\begin{eqnarray}}
\def\eea{\end{eqnarray}}

\begin{document}
\vspace*{4cm}
\title{Dark Energy Phenomenology}  % max 4 pages for me

\author{ Martin Kunz }

\address{Astronomy Centre, University of Sussex, Falmer, Brighton BN1 9QH, UK}

\author{ Luca Amendola }

\address{INAF/Osservatorio Astronomico di Roma, Via Frascati 33, 00040 Monteporzio
Catone, Roma, Italy}

\author{ Domenico Sapone }

\address{Département de Physique Théorique, Université de Genève, 24 quai
E. Ansermet, 1211 Genève 4, Switzerland}

\maketitle\abstracts{
We discuss the phenomenology of the dark energy in first order
perturbation theory, demonstrating that the dark energy cannot be
fully constrained unless the dark matter is found, and that there
are two functions that characterise the observational properties
of the dark sector for cosmological probes. We argue that measuring
these two functions should be an important goal for observational
cosmology in the next decades.
}

\section{Introduction}

The observed accelerated expansion of the Universe is considered as the main
mystery in modern cosmology and one of the major issues confronting theoretical
physics at the beginning of the new Millennium. Although a plethora of models
have been constructed, none of them is really appealing on theoretical grounds.
An alternative approach in this situation is to construct general parametrisations
of the dark energy, in the hope that measuring these parameters will give us some
insight into the mechanism underlying the dark energy phenomenon.

A successful example for such a phenomenological parametrisation
in the dark energy context is the equation
of state parameter of the dark energy component, $w\equiv p/\rho$. If
we can consider the Universe as evolving like a homogeneous and isotropic 
Friedmann-Lemaitre-Robertson-Walker (FLRW) universe, and if the dark energy
is not coupled to anything except through gravity, then $w(z)$ completely
specifies its evolution: The dark energy (or anything else)
is described by the homogeneous energy density $\rho_\dde$ and the isotropic
pressure $p_\dde$, corresponding to the $T_0^0$ and $T_i^i$ elements respectively
in the
energy momentum tensor in the rest-frame of the dark energy. Any other non-zero
components would require us to go beyond the FLRW description of the Universe.
The evolution of $\rho$ is then governed by the "covariant conservation" equation
$T_{\mu\, ;\nu}^\nu=0$ which is just
\beq
\dot{\rho}_\dde = -3 H (\rho_\dde+p_\dde) = -3 H (1+w) \rho_\dde .
\eeq
As long as $H\neq0$ we can choose the evolution of $\rho_\dde$ through the choice
of $p_\dde$ or equivalently the choice of $w$. It is usually more convenient to
quote $w$ since it is independent of the absolute scale of $\rho$ and $p$ and many
fluids give rise to simple expressions for $w$, for example $w=0$ for pressureless
dust, $w=1/3$ for a radiation fluid, or $w=-1$ for a cosmological
constant.

The only observationally accessible quantity is the expansion rate of the universe
$H$, given by the Friedmann equation,
\beq
H^2 = \left(\frac{\dot{a}}{a}\right)^2 = \frac{8\pi G}{3} \left( \rho_m + \rho_{\rm DE} \right) .
\eeq
For simplicity we neglect radiation and assume that space is flat. The distances
are integrals of $1/H$, and $H$ can be directly measured with some methods like
baryonic acoustic oscillations (BAO) \cite{bao} or the dipole of the supernova 
distribution \cite{sndipole}.
The relative abundance of matter today $\Omega_m$, could also be measured, but at
the moment this is not possible directly, as we have not yet been able to detect
the dark matter in experiments. Thus, at this level (often called the "background"
level) only $H$ is measureable, and we would like to infer $w$, $\Omega_m$ and
$\Omega_\dde = 1-\Omega_m$.

\section{The dark sector}

Assuming that we have a perfect measurement of $H$, we can then directly derive
an expression for $w$:

\beq
w(z) = \frac{H(z)^2-\frac{2}{3}H(z)H'(z)(1+z)}{H_0^2 \Omega_m (1+z)^3 - H(z)^2}
\eeq

This expression exposes an awkward problem with our data: since we do not know
$\Omega_m$, we find a solution for $w$ for {\em every} choice of $\Omega_m$. We
can therefore not measure both $w$ and $\Omega_m$ simultaneously, without approximations.
Consequently, we are left with a one-parameter family of possible $w$'s \cite{mk}.

This may surprise the reader, since she or he may have thought
that $\Omega_m\approx 0.25$ from observations. However, this conclusion
requires assumptions on the nature of the dark energy, and so let us spend
a few lines looking at them. One possible assumption is to impose
$w=-1$, ie to demand that the dark energy be a cosmological constant. Alternatively,
we would reach similar limits on $\Omega_m$ by being slightly more general and
allowing $w$ to be a free constant. Both of these are very strong assumptions
if we wanted to actually learn something about the dark energy from the data!
Indeed, we should conclude that any dark energy analysis that uses only data 
based on measurements of "background" quantities derivable from $H$ like 
distances or ages {\em must} find no constraint on $\Omega_m$, or else the
parametrisation of $w$ is not sufficiently general.\footnote{Adding even
more freedom to the dark energy, like allowing for couplings to the dark matter,
makes the degeneracy even worse.}

Often more data is included the analysis, for example the angular power spectrum
of temperatures anisotropies in the cosmic microwave background (CMB). This data
not only constrains the expansion history of the Universe, but also the clustering
of the fluids. Under the additional (and also strong) assumption that the dark
energy does not cluster significantly, we are then able to separate the dark 
matter and the dark energy in this case \cite{mk}.

To include the CMB data we need to improve our description of the universe, by
using perturbation theory. If we work in the Newtonian gauge, we add two gravitational
potentials $\phi$ and $\psi$ to the metric. They can be considered as being
similar to $H$ in that they enter the description of the space-time (but they
are functions of scale as well as time). Also the energy momentum tensor becomes
more general, and $\rho$ is complemented by perturbations $\delta\rho$ as well
as a velocity $V_i$. The pressure $p$ now can also have perturbations $\delta p$
and there can further be an anisotropic stress $\pi$.

The reason why we grouped the new parameters in this way is to emphasize their
role: at the background level, the evolution of the universe is described by $H$,
which is linked to $\rho$ by the Einstein equations, and $p$ controls the evolution
of $\rho$ but is a priori a free quantity describing the physical properties of the 
fluid. Now in addition there are $\phi$ and $\psi$ describing the Universe, and
they are linked to $\delta\rho_i$ and $V_i$ of the fluids through the Einstein 
equations. $\delta p_i$ and
$\pi_i$ in turn describe the fluids. Actually, there is a simplification: the
total anisotropic stress $\pi$ directly controls the difference between the potentials,
$\phi-\psi$.

\section{Dark energy phenomenology}

We have seen in the previous section that a general dark energy component can be
described by phenomenological parameters like $w$, even at the level of first
order perturbation theory. This description adds two new paramters $\delta p$
and $\pi$, which are both functions of scale as well as time. These parameters
fully describe the dark energy fluid, and they can in principle be measured.

However, recently much interest has arisen in modifying GR itself to explain
the accelerated expansion without a dark energy fluid. What happens if we
try to reconstruct our parameters in this case? Is it possible at all?

Let us assume that the (dark) matter is three-dimensional and conserved, and
that it does not have any direct interactions beyond gravity. We assume further
that it and the photons move on geodesics of the same (possibly effective) 
3+1 dimensional
space-time metric. In this case we can write the modified Einstein equations as
\beq
X_{\mu\nu} = -8\pi G T_{\mu\nu}
\eeq
where the matter energy momentum tensor still obeys $T_{\mu\, ;\nu}^\nu=0$. While
in GR this is a consequence of the Bianchi identies, this is now no longer the
case and so this is an additional condition on the behaviour of the matter.

In this case, we can construct $Y_{\mu\nu} = X_{\mu\nu} - G_{\mu\nu}$,
so that $G_{\mu\nu}$ is the Einstein tensor of the 3+1 dimensional
space-time metric and we have that
\beq
G_{\mu\nu} = -8\pi G T_{\mu\nu} - Y_{\mu\nu} ,
\eeq
Up to the prefactor we can consider $Y$ to be the energy momentum
tensor of a dark energy component. This component is also covariantly conserved
since $T$ is and since $G$ obeys the Bianchi identities. The equations governing
the matter are going to be exactly the same, by construction, so that the
effective dark energy described by $Y$ mimics the modified gravity model \cite{hu}.

By looking at $Y$ we can then for example extract an effective anisotropic
stress and an effective pressure perturbation and build a dark energy model
which mimics the modified gravity model and leads to {\em exactly} the
same observational properties \cite{ks}.

This is both good and bad. It is bad since cosmology cannot directly distinguish
dark energy from modified gravity\footnote{Although there could be clear hints, e.g.
a large anisotropic stress would favour modified gravity since in these
models it occurs generically while scalar fields have $\pi_i=0$.}. However, it is good since there is a clear
target for future experiments: Their job is to measure the two additional functions
describing $Y$ as precisely as possible.

\section{Forecasts for future experiments}

\begin{figure}[ht]
\begin{center}
\epsfig{figure=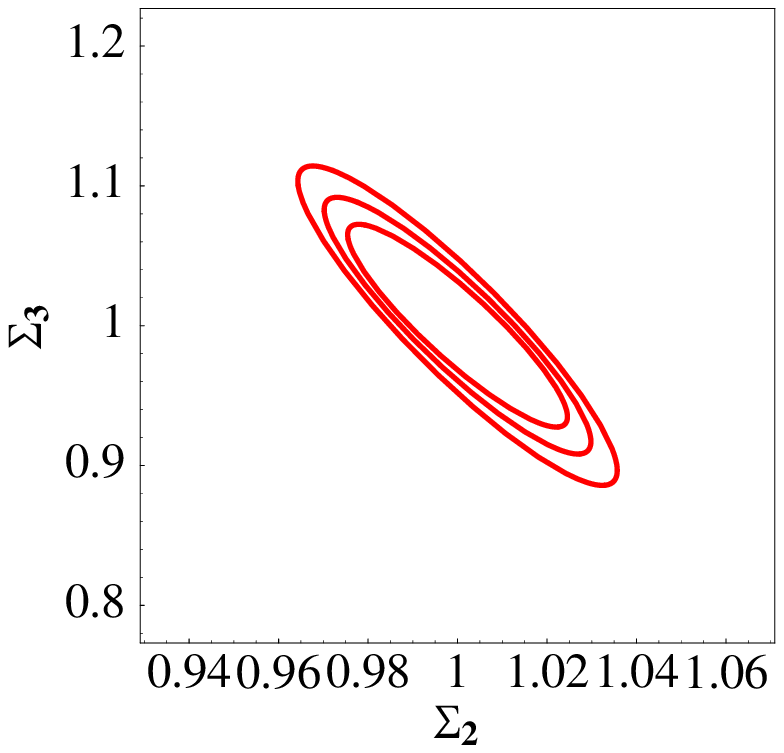,height=1.7in}
\epsfig{figure=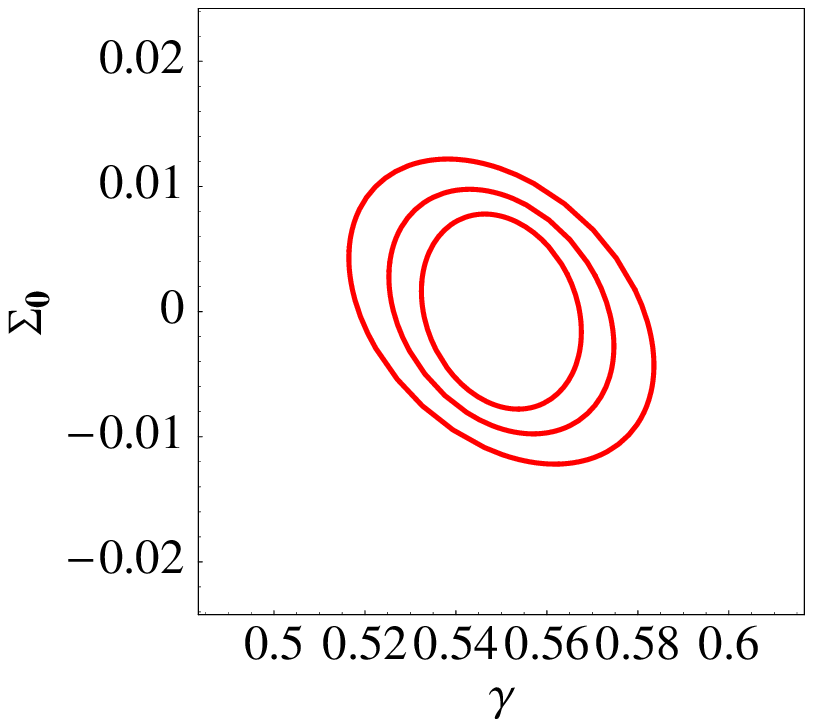,height=1.8in}
\end{center}
\caption{Forecast of a how a future weak lensing survey could constrain
the behaviour of the dark energy or modified gravity at the perturbation
level.
\label{fig:radish}}
\end{figure}

As an example, we show in Fig.~\ref{fig:radish} 
two graphs from \cite{aks}. Here a different choice was
made for the parametrisation of the extra dark energy freedom: The logarithmic
derivative of the dark matter perturbations were characterised by 
$\Omega_m(z)^\gamma$, and the deviations of the lensing potential $\phi+\psi$
from a fiducial cosmology with unclustered dark energy (chosen arbitrarily as a 
reference point) with a function $\Sigma$. Because of space constraints, 
we refer the reader to \cite{aks} for more details.

\section{Conclusions}

We have shown in these proceedings that both dark energy and modified gravity
cosmologies can be described at the level of first-order perturbation theory
by adding two functions to the equation of state parameter $w(z)$. This allows
to construct a phenomenological parametrisation for the analysis of e.g. CMB
data, weak lensing surveys or galaxy surveys, which depend in essential ways
on the behaviour of the perturbations.

Different choices are possible for these two functions. For example, one
can directly use the gravitational potentials $\phi(k,t)$ and $\psi(k,t)$
and so describe the metric\footnote{This would correspond to using $H(z)$ rather
than $w(z)$ to describe the dark energy at the background level.}. Alternatively 
one can use the parameters which describe the dark energy, the pressure perturbation
$\delta p(k,t)$ and the anisotropic stress $\pi(k,t)$. The pressure perturbation
can be replaced by a sound speed $c_s^2$, which then has to be allowed to depend
on scale and time. As argued above, these parameters can {\em also} describe
modified gravity models, in which case they do of course not have a physical reality.
Also other choices are possible, the important message is that {\em only two new
functions} are required (although they are functions of scale $k$ and time $t$).
Together with $w(z)$, they span the complete model space for both modified gravity 
and dark energy models in
the cosmological context (ie without direct couplings, and for 3+1 dimensional matter
and radiation moving on geodesics of a single metric). By measuring them, we extract
the {\em full information} from cosmological data sets to first order.

These phenomenological functions are useful in different contexts. Firstly they can
be used to analyse data sets and to look for {\em general} departures from e.g.
a scalar field dark energy model. If measured, they can also give clues to the physical
nature of whatever makes the expansion of the Universe accelerate. Finally, they
are useful to forecast the performance of future experiments in e.g. allowing to
rule out scalar field dark energy, since for this explicit alternatives are needed.

%\section*{Acknowledgments}
%Part of this work was supported by the Swiss National Science Foundation.

\end{document}